\documentclass{iau}   
\usepackage{graphicx} 

\newcommand{\OI}{O\,{\sc i}}  

\newcommand{\CII}{C\,{\sc ii}}
\newcommand{\HII}{H\,{\sc ii}}
\newcommand{\HI}{H\,{\sc i}}

\title[The ALMA view of PDRs] 
{The ALMA view of UV-irradiated cloud edges: 
unexpected structures and processes}

\author[Javier R. Goicoechea et al.]   
{Javier R. Goicoechea$^1$, S. Cuadrado$^1$, J. Pety$^{2,3}$, A. Aguado$^4$, 
J.~H.~Black$^5$, E.~Bron$^1$, 
J.~Cernicharo$^1$, E.~Chapillon$^{2,6}$, A. Fuente$^7$,\\ M.~Gerin$^3$, C. Joblin$^8$ 
 \and O. Roncero$^9$, B. Tercero$^7$}
 
\affiliation{$^1$ICMM-CSIC, 
Calle Sor Juana Ines de la Cruz 3, 28049 Cantoblanco, Madrid, Spain.\\  
$^2$IRAM, 300 rue de la Piscine, 38406 Saint-Martin-d’H\`eres, France.\\
$^3$LERMA, Observatoire de Paris, PSL Research University, CNRS, Sorbonne Universit\'es, 
UPMC Univ. Paris 06, ENS, 75005 Paris, France\\
$^4$Facultad de Ciencias, Universidad Aut\'onoma de Madrid, 28049 Madrid, Spain\\
$^5$Chalmers University of Technology, Onsala Space Observatory, 43992 Onsala, Sweden\\
$^6$OASU/LAB-UMR5804, CNRS, Universit\'e Bordeaux, 33615 Pessac, France\\
$^7$Observatorio Astron\'omico Nacional (IGN). Apartado 112, 28803 Alcal\'a 
de Henares, Spain\\
$^8$Universit\'e de Toulouse, UPS-OMP, 
CNRS, Institut de Recherche en Astrophysique et Plan\'etologie , 9 Avenue du Colonel Roche, BP 44346, 31028 Toulouse, France.\\
$^9$Instituto de F\'{\i}sica Fundamental (IFF-CSIC), Calle Serrano 123, 28006 Madrid, Spain.}

\pubyear{2017}
\volume{IAUS 332}  
\jname{Astrochemistry VII – Through the Cosmos from Galaxies to Planets}
\editors{M. Cunningham, T. Millar \& Y. Aikawa, eds.}
\begin{document}

\maketitle

\begin{abstract}
Far-UV photons (FUV, $E < 13.6$\,eV) from hot massive stars regulate, or at least influence, the heating, ionization, and chemistry of most of the neutral interstellar medium (\HI~and H$_2$ clouds). Investigating the interaction between FUV radiation and interstellar matter (molecules, atoms and grains) thus plays an important role in astrochemistry.

The Orion Bar, an interface region between the Orion~A molecular cloud and the 
\HII\, region around the Trapezium cluster, is a textbook example of a strongly illuminated dense PDR (photodissociation region).  The Bar is illuminated by a  FUV field of a few 10$^4$ times the mean interstellar radiation field. Because of its proximity and nearly edge-on orientation, it provides a very good template to investigate the chemical content, structure, and dynamics of a strongly 
irradiated molecular cloud edge.  We have used ALMA to mosaic a small field of the Bar where the critical transition from atomic to molecular gas takes place. These observations provide an unprecedented sharp view of this transition layer ($\lesssim 1''$ resolution or $\lesssim 414$\,AU). The resulting images (so far in the rotational emission of CO, HCO$^+$,
H$^{13}$CO$^+$, SO$^+$, SO, and reactive ions SH$^+$ and HOC$^+$) show the small-scale structure in gas density and temperature, and  the steep  abundance gradients.  The  images reveal a pattern of high-density substructures, photo-ablative gas flows and instabilities at the edge of the molecular cloud. These first ALMA images thus show a more complex morphology than the classical clump/interclump static model of a PDR.

In order to quantify the chemical content in strongly FUV-irradiated gas, we have  also
used the IRAM-30\,m telescope to carry out a complete line-survey of the illuminated edge of the Bar in the millimeter domain. Our observations reveal the presence of complex organic molecules (and precursors) that were not expected in such a harsh environment. In particular, we have  reported the first detection of the unstable \textit{cis} conformer of formic acid (HCOOH) in the ISM. The energy barrier to internal rotation (the conversion from \textit{trans} to \textit{cis}) is approximately 4827\,cm$^{-1}$ ($\approx7000$\,K). 
Hence, its detection is surprising. The low inferred \textit{trans}-to-\textit{cis} abundance ratio of 2.8$\pm$1.0  supports a photoswitching mechanism: a given conformer absorbs a FUV stellar photon that radiatively excites the molecule to electronic states above the interconversion barrier. Subsequent fluorescent decay leaves the molecule in a different conformer form. This mechanism, which we have specifically studied with \textit{ab initio} quantum calculations, was not considered so far in astrochemistry although it can affect the structure of a variety of molecules in PDRs.

\end{abstract}

\firstsection 
\section{Introduction: the role of far-UV stellar  radiation on the ISM}

Far-UV photons (FUV: $E <13.6$\,eV) from hot massive O and B stars regulate, or at least greatly influence, the dynamics, heating, ionization, and chemistry of the neutral  interstellar medium (ISM). From a global perspective, the photo-induced processes that take place in FUV-illuminated environments reflect the radiative feedback from such stars. This feedback takes place at very different spatial scales: in star-forming regions of the Milky Way, the ISM of starburst galaxies, and back in cosmic history: see the rapidly increasing number of ALMA detections of the [\CII]\,158\,$\mu$m fine-structure line at \mbox{high-redshifts} (\cite[e.g., Capak et al. 2015, \textit{Nature})]{Capak_2015}. This far-IR emission line of  ionized carbon (C$^+$) was first detected in the 80s (\cite[Russell et al. 1980]{Russell_1980}) and is often the brightest and most important line coolant of the neutral ISM (hydrogen atoms in neutral form).

        Investigating the interaction between FUV radiation and interstellar baryonic matter (atoms, molecules and dust grains) ultimately contributes to a better understanding of a varity of FUV-irradiated objects and environments of the galaxy: the diffuse interstellar clouds, the interfaces between 
\HII~regions near young massive stars and their natal clouds, reflection and planetary nebulae around evolved stars, and the externally irradiated protoplanetary disks (proplyds). All these environments in which the underlying physics and chemistry are driven by the presence of FUV photons are generically defined as photodissociation regions (PDRs; see the review paper by \cite[Hollenbach \& Tielens 1999]{Hollenbach_1999}). FUV photons with wavelengths longer than 911\,\AA~do not ionize H atoms (as in \HII~regions) but do dissociate molecules and ionize atoms with ionization potental below 13.6\,eV (C, S, Si, Fe, P, etc.). Hence, PDRs are essentialy neutral, with ionization fractions as high as [$e^-$]/[H]$\simeq$10$^{-4}$ when most of the electrons are produced by the ionization of carbon atoms, a key process in the ISM (\cite[see e.g., Dalgarno \& McCray 1972]{Dalgarno_1972}).

\begin{figure}[b]
\begin{center}
 \includegraphics[width=5.3in]{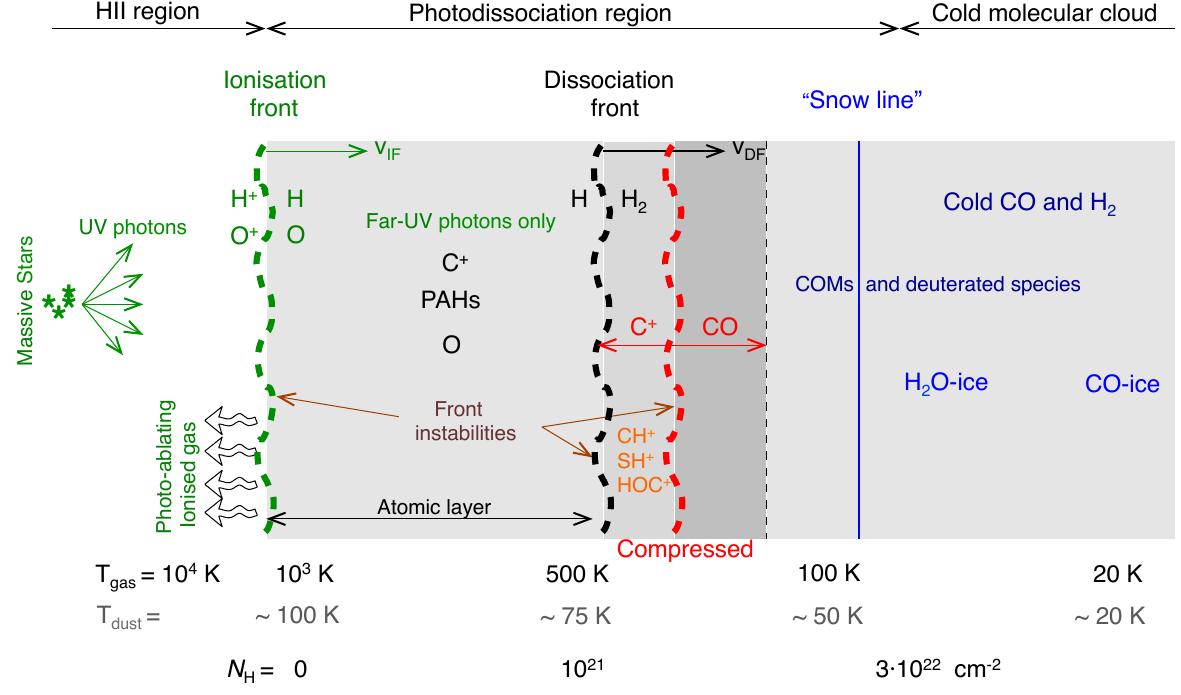} 
 \caption{Structure of a strongly FUV-irradiated  cloud edge (from \cite[Goicoechea et al. 2016]{Goico_2016}).}
   \label{fig1}
\end{center}
\end{figure}

         PDRs are very relevant because they host the critical conversion from atomic to molecular ISM (the H$^+$/H/H$_2$ and C$^+$/C/CO transition zones, see  Figure\,\ref{fig1}). This stratification occurs as the column density of gas and dust increases deeper inside clouds and the FUV photon flux is attenuated
(see \cite[Goicoechea \& Le Bourlot 2007]{Goicoechea_2007} and references therein).
    In addition, PDRs emit most of the IR radiation from the ISM of star-forming galaxies: mid-IR bands from polycyclic aromatic hydrocarbons (PAHs), IR H$_2$ ro-vibrational lines,  [\CII]\,158\,$\mu$m and [\OI]\,63\,$\mu$m fine-structure lines, and warm dust continuum: after absorbing a FUV photon, grains are heated and radiation is reemitted in the mid- and far-IR. At the same time, photoelectrons ejected from grains and PAHs heat the gas.
    
         The theoretical study and first thermo-chemical models of dense PDRs started to strongly develop more than 30 years ago. 
         With the above broad definition, PDRs represent a significant fraction of the neutral atomic and molecular material in the Milky Way, perhaps more than 90\%: all the gas and dust in the ISM between $A_V$$\approx$0.1 and 7\,mag (\cite[Hollenbach \& Tielens 1999]{Hollenbach_1999}). At deeper cloud depths, the flux of stellar FUV photons is almost completely attenuated, gas-phase molecules and atoms freeze, and a much slower chemistry and weak ionization take place driven by  cosmic-ray particles (see Figure\,\ref{fig1}).

        In addition to being excellent tracers of the gas physical conditions (density, temperature, FUV field strenght, etc.),  PDR observational diagnostics ([\CII]\,158\,$\mu$m, PAH bands,  H$_2$ lines, and far-IR continuum emission) can be used to estimate the star-formation rate in galaxies. More locally, PDR diagnostics can be used to determine the radiative feedback of massive stars on their natal clouds. Understanding how molecular clouds are evaporated by a strong FUV radiation field is important to determine their lifetimes, and how the star-formation process can be quenched in a given region. On the other hand, the dynamical effects induced by a strong stellar FUV field (e.g. radiation pressure), by the winds of massive stars, or by Supernova explosions could trigger, even regulate, the formation of a new generation of low-mass stars (\cite[Hollenbach \& Tielens 1999]{Hollenbach_1999}).
        
   In this contributed paper we summarize our latest results regarding the study of the Orion Bar PDR, an excellent astrochemical laboratory to investigate the interaction of interstellar matter with a strong flux of FUV photons from nearby massive stars.\\

\begin{figure}[t]
\begin{center}
 \includegraphics[width=4.1in]{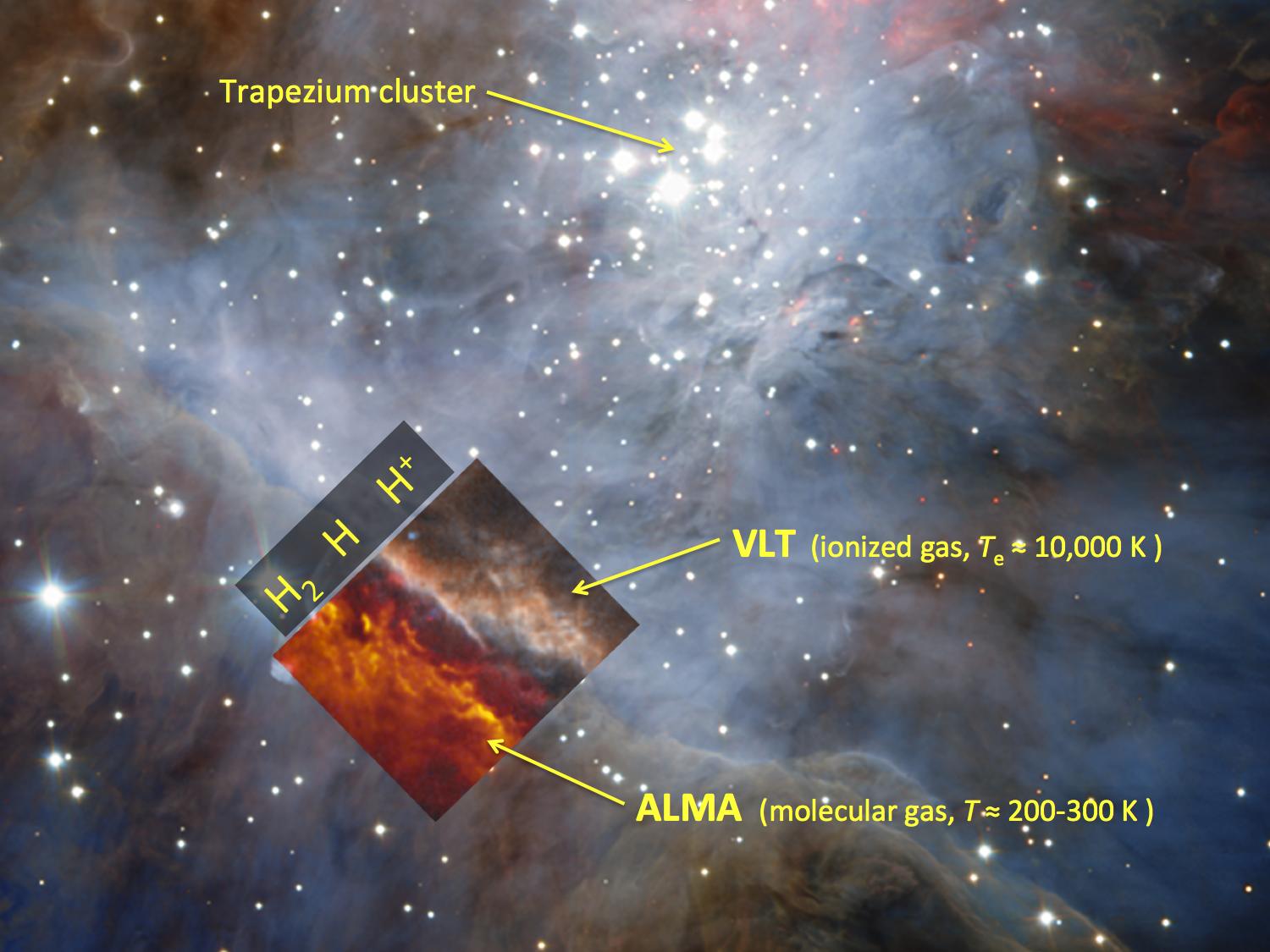} 
 \caption{Part of the Orion nebula (ionized gas irradiated by UV photons from the Trapezium cluster) observed by the VLT, and a small field of the Orion Bar PDR observed by ALMA in molecular gas emission (adapted from \cite[Goicoechea et al. 2016]{Goico_2016}, \textit{Nature}).}
   \label{fig2}
\end{center}
\end{figure}

\textbf{The Orion Bar}: Owing to its nearly edge-on orientation, ``the Bar'' (an interface region between the Orion A molecular cloud and the \HII~region around the Trapezium cluster, see Figure\,\ref{fig2}) is a prototypical strongly irradiated PDR with $G_0$$\simeq$2$\times$10$^4$ ($G_0$=1.7 is roughly the mean interstellar FUV field in Habing units, equivalent to a few 10$^8$~photons\,cm$^{-2}$\,s$^{-1}$ between 13.6 and 6\,eV). Its closeness ($\sim$414 pc), high gas temperatures ($T_{\rm k}$$\simeq$150-300\,K) and thus bright molecular lines, makes it an ideal target for high signal-to-noise spectral-imaging at different wavelengths. 
     In the last years, our team has led several efforts to constrain the molecular content of the Orion Bar by using the \textit{Herschel Space Observatory} and the  IRAM-30\,m telescope, and more recently to study the small-scale density and temperature structures with ALMA. From these works \textbf{we highlight}:\\
\textbf{(i)} A line survey of the Bar edge in the 3 to 0.8\,mm bands, complemented with $\sim$2$'$$\times$2$'$  maps at $\sim$7$''$ resolution  with the IRAM-30m telescope (\cite[Cuadrado et al. 2015]{Cuadrado_2015},  \cite[2016]{Cuadrado_2016}, \cite[2017]{Cuadrado_2017}). Once thought to be a harsh environment to develop a rich chemistry, our observations show that PDRs host unique chemical features: reactive ions (SH$^+$ or HOC$^+$; \cite[Fuente et al. 2003]{Fuente_2003}, \cite[Goicoechea et al. 2017]{Goico_2017}), hydrocarbon ions (C$_3$H$^+$; \cite[Pety et al. 2012]{Pety_2012}; 
\cite[Cuadrado et al. 2015]{Cuadrado_2015}), and unstable isomers of organic species (\textit{cis}-HCOOH, \cite[Cuadrado et al. 2016]{Cuadrado_2016}) that are not seen, or that are not abundant, in regions shielded from FUV radiation.\\
\textbf{(ii)} \textit{Herschel}/HIFI  large-scale (7.5$'$$\times$11.5$'$) velocity-resolved maps of Orion~A (including the Bar) in FUV-radiation diagnostics such as [\CII]\,158\,$\mu$m (\cite[Goicoechea et al. 2015]{Goico_2015}).\\
  \textbf{(iii)} Pilot ALMA mosaics of a small field ($\sim$50$''$$\times$50$''$) of the Bar in several molecular line diagnostics at $\sim$1$''$ resolution (Figures\,\ref{fig3}
and \ref{fig4}) (\cite[Goicoechea et al. 2016]{Goico_2016}, \cite[2017]{Goico_2017}).\\ 

\section{Morphology of the PDR: small-scale structures and flows}

\begin{figure}[th]
\begin{center}
 \includegraphics[width=5.3in]{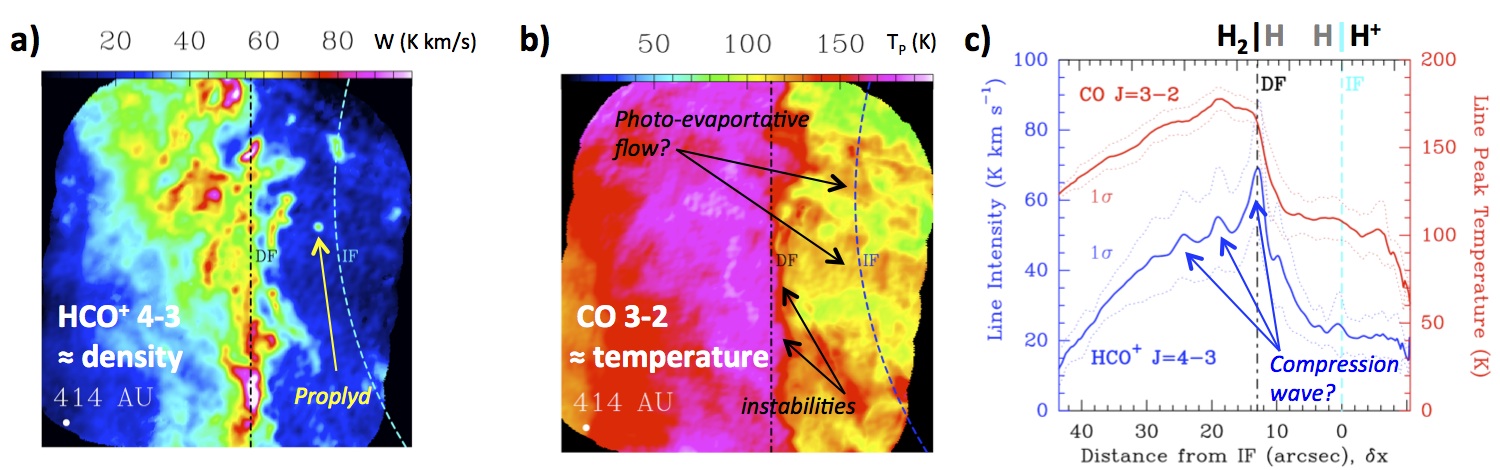} 
 \caption{ALMA view of a small field of the Orion Bar in
two tracers of the molecular gas: a) HCO$^+$ $J$=4-3 and b) CO $J$=3-2. 
c) Vertically averaged intensity cuts perpendicular to the Orion Bar. The flux of FUV
photons decreases from right to left in these rotated images.}
\label{fig3}
\end{center}
\end{figure}

Surprisingly, our pilot ALMA images show that there is no appreciable offset between the peak of the H$_2$ vibrational emission at 2.2\,$\mu$m (delineating  the H$_2$ dissociation front)  and the edge of the observed CO and HCO$^+$ rotational emission (Figures\,\ref{fig2} and \ref{fig3}). This implies that the H/H$_2$ and C$^+$/C/CO transition zones are very close, much closer than stationary PDR model predictions. In addition, our ALMA mosaics reveal a fragmented ridge of high-density substructures, photoablative gas flows and instabilities at the irradiated molecular cloud surface (Figure\,\ref{fig3}; see \cite[Goicoechea et al. 2016]{Goico_2016} for details). These results suggest that the cloud edge has been compressed by a high-pressure wave that is moving into the molecular cloud, demonstrating that dynamical effects are  more important than previously thought.
     Whether the newly detected small-scale structures could be the seed of future star-forming clumps (e.g., by merging into more massive collapsing clumps) is uncertain and is subject of study. Gravitational collapse is not yet apparent from their density distribution nor from their low masses. However, our images resolve the 
     \mbox{HCO$^+$ $J$=4-3} emission from \textit{proplyd~203-506} (Figure\,\ref{fig3}a and
\cite[Champion et al. 2017]{Champion_2017}). Interestingly, its location in the atomic PDR ($A_V$=0 to 1\,mag,  thus in-between the ionization and dissociation fronts) suggests that this proplyd (a low-mass protostar and protoplanetary disk) could have emerged from the molecular cloud after surviving the passage of the dissociation front. Even on its early cycles (the antena array was not yet completed at the time of our observations) ALMA clearly goes a step beyond in our understanding of the closest, and most iconic, massive star forming region. The results obtained from these pilot ALMA mosaics were unexpected and challenge many years of PDR modelling.
However, they were obtained by imaging a small fraction of the  Bar ($\approx$10-20\% of its size; see Figure\,\ref{fig2}). In order to support and better understand our findings, it is mandatory to obtain ALMA images of the entire Bar, and of course, to observe other (more distant) massive star forming regions, i.e., at different stages of evolution and having different stellar populations, thus different number of ionizing massive stars.

\section{The short, but interesting, life of reactive ions}

Reactive  ions (CH$^+$, SH$^+$, CO$^+$, HOC$^+$, ...)  are transient species for which the timescale of reactive collisions with H$_2$, H, or $e^-$ (leading to a chemical reaction, and thus molecule destruction) is comparable to, or shorter than, that of inelastic collisions (\cite[Black 1998]{Black_1998}).
The formation of  reactive ions such as SH$^+$, for example, 
 depends on the availability of S$^+$ ions (i.e., on the availability  of FUV photons that can ionize sulphur atoms) and on the presence of excited H$_2$ (either \mbox{UV-pumped} or hot and thermally excited). 
These two ingredients allow overcoming the high endothermicity (and sometimes energy barrier) of some of the key initiating chemical reactions 
(\cite[Gerin et al. 2016]{Gerin_2016} for a review).
The reaction \mbox{S$^+$ + H$_2$(v) $\rightarrow$ SH$^+$ + H}, in particular, is
endothermic by \mbox{$\Delta E/k\simeq 9860$\,K} if v=0, and only becomes exothermic when v$\geq$2 (\cite[Zanchet et al. 2013]{Zanchet_2013}).
Despite their short lifetimes, from a few years in diffuse gas to only a few hours (!) in  dense PDR gas, reactive ions can be detected and probe energetic processes in FUV-irradiated warm molecular gas.

\begin{figure}[t]
\begin{center}
 \includegraphics[width=4.9in]{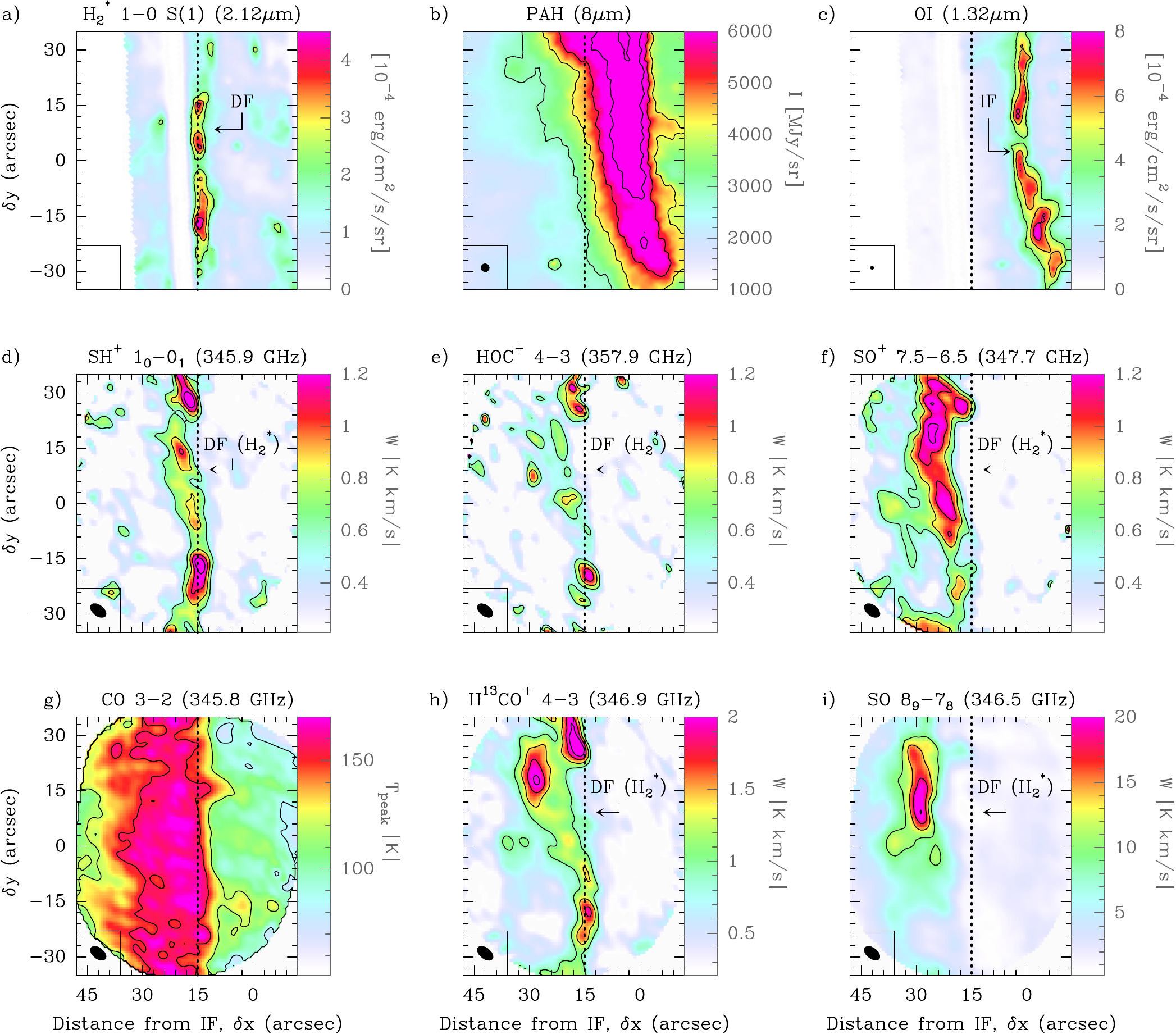} 
 \caption{ALMA-ACA observations of several molecules, including the reactive ions SH$^+$ and HOC$^+$ and complementary images of the Bar
 (\cite[Goicoechea et al. 2017]{Goico_2017}). All images have been rotated to bring the FUV illuminating direction in the horizontal direction (from the right).  The upper row shows images of a) the H$_2$ $v$=1-0 $S$(1) line at 2.12\,$\mu$m, delineating the dissociation front (DF) 
 (\cite[Walmsley et al. 2000]{Walmsley_2000}); b) the Spitzer 8\,$\mu$m emission produced mainly by PAHs, and c) the fluorescent \OI~1.32$\mu$m~line arising from the \HII/PDR boundary (\cite[Walmsley et al. 2000]{Walmsley_2000}).}
\label{fig4}
\end{center}
\end{figure}

We have used the ALMA-ACA array to obtain high angular resolution (5$''$$\times$3$''$) images of some reactive ions toward the  Bar (\cite[Goicoechea et al. 2017]{Goico_2017}). The observed SH$^+$ and HOC$^+$ emission is restricted to a narrow layer of 2$''$- to 10$''$-width \mbox{($\approx$800 to 4000\,AU} depending on the assumed PDR geometry) that follows the vibrationally excited H$_{2}^{*}$ emission (Figure\,\ref{fig4}). Both ions efficiently
  form very close to the \mbox{H/H$_2$} transition zone,
  at a depth of $A_{\rm V}\lesssim1$\,mag into
the neutral cloud, where abundant C$^+$, S$^+$, and H$_{2}^{*}$ coexist. The observed  ions have low rotational temperatures   
(\mbox{$T_{\rm rot}$$\approx$10-30\,K$\ll$$T_{\rm k}$}) and narrow line-widths  \mbox{($\sim$2-3\,km\,s$^{-1}$)}, a factor of $\simeq$2 narrower that those of the lighter  reactive ion CH$^+$ (\cite[Nagy et al. 2013]{Nagy_2013}). In
\cite[Goicoechea et al. (2017)]{Goico_2017} we show that this 
is consistent with the higher 
reactivity\footnote{For the physical  conditions at the edge of the 
 Bar, the destruction time scales ($\tau_{\rm D}$) of CH$^+$, HOC$^+$, and SH$^+$ are $\simeq$5\,h, $\simeq$25\,h and $\simeq$50\,h respectively.
Comparing with the timescales for collisional and
radiative excitation shows 
that  CH$^+$ molecules are excited by radiation many times during their short lifetime,
but not by collisions. 
Hence, CH$^+$ can remain rotationally warm (owing to formation pumping;
\cite[Nagy et al. 2013]{Nagy_2013}, \cite[Godard \& Cernicharo 2013]{Godard_2013}) while it emits. 
On the other hand, comparing \mbox{$\tau_{\rm D}$(HOC$^+$)} 
with the timescales for collisional and radiative excitation
shows that HOC$^+$ molecules (and likely the other heavier ions as well) are excited by collisions several times during their lifetime. Inelastic collisions thus can drive their rotational populations to lower $T_{\rm rot}$ relatively fast 
($T_{\rm rot}$(HOC$^+$)\,$\ll$\,$T_{\rm rot}$(CH$^+$)$\approx$150\,K; \cite[Goicoechea et al. 2017]{Goico_2017}).}
 and faster radiative pumping rates of  CH$^+$  compared to the heavier ions, which are driven relatively more  quickly toward smaller velocity dispersion by elastic collisions and toward lower $T_{\rm rot}$ by inelastic collisions (see also \cite[Black 1998]{Black_1998}). 
Regardless of the excitation details, the ALMA-ACA images show that SH$^+$ and HOC$^+$  
clearly trace the most exposed layers of the FUV-irradiated molecular cloud surface.

\section{Yes, COMs are seen in strongly FUV-irradiated gas}

Observations towards  UV-shielded cold cores (e.g.~TMC\,1, L1689B, or B1-b) have revealed molecules once considered to be present only in hot cores and corinos 
(\cite[e.g., Cernicharo et al. 2012]{Cerni_2012}). In these environments, complex organic molecules (COMs) are thought to form on the surface of grains and to be released  through non-thermal desorption processes, chemical desorption, direct desorption by cosmic rays impacts, or by secondary FUV-photon induced processes.
Until recently, observational studies of  environments illuminated  by stellar FUV photons were more scarce. 
Guzm\'an et al. (\cite[2014]{Vivi_2014}) presented the  unexpected detection of \textit{trans}-HCOOH, CH$_{2}$CO, CH$_{3}$CN, CH$_{3}$OH, CH$_{3}$CHO, and CH$_{3}$CCH at the edge of the Horsehead (a low \mbox{FUV-illumination} PDR with \mbox{$G_0$$\approx$100}). They found enhanced abundances compared to a nearby cold and dense core shieled from external FUV radiation. 
Guzm\'an et al. (\cite[2014]{Vivi_2014}) proposed that owing to the cold dust temperatures in this PDR ($T_{\rm d}$$\lesssim$30\,K, i.e., grains should be coated by ices), ice-mantle photodesorption processes dominate the formation of gaseous COMs. In lower density translucent clouds ($G_0$$\approx$1),
only H$_2$CO has been unambiguously detected (\cite[Liszt et al. 2006]{Liszt_2006}). These observations might suggest that the presence of COMs diminishes  as $G_0$/$n_{\rm H}$ increases, and hence, that COMs might not be present in strongly FUV-irradiated gas.

Despite being a much harsher environment, we  detected more than 250 lines from COMs and related precursors toward the irradiated edge of the Orion Bar: H$_{2}$CO, CH$_{3}$OH, HCO, H$_{2}$CCO, CH$_{3}$CHO, H$_{2}$CS, HCOOH, CH$_{3}$CN, CH$_{2}$NH, HNCO, H$_{2}^{13}$CO, and HC$_{3}$N (in decreasing order of abundance, \cite[Cuadrado et al. 2017]{Cuadrado_2017}).
Taking into account the elevated gas ($\approx$200\,K) and dust ($\approx$60\,K) temperatures, we suggested the following scenarios for the formation of COMs in the Orion Bar:
\textit{(i)} hot gas-phase reactions not included in current chemical models;
\textit{(ii)} bare (little ice) warm grain surface chemistry; or \textit{(iii)} the PDR dynamics is such that COMs previously formed in cold icy grains deeper inside the
molecular cloud desorb and advect into the PDR  (\cite[Cuadrado et al. 2017]{Cuadrado_2017}).

\subsection{\textit{Trans-cis} photoisomerization of HCOOH}

Conformational isomerism refers to isomers (molecules with the same formula but
different chemical structure) having the same chemical bonds but different geometrical
orientations around a single bond. Such isomers are called conformers. An energy barrier often limits the isomerization. This barrier can be overcome by light. Photoisomerization (or photoswitching) has been studied in ice IR-irradiation experiments, in biological processes, and, for large polyatomic molecules, in gas-phase experiments. HCOOH is the simplest organic acid and has two conformers (\textit{trans} and \textit{cis}) depending on the orientation of the
hydrogen single bond. The most stable \textit{trans} conformer was the first acid detected in the ISM. Gaseous \textit{trans}-HCOOH shows moderate abundances towards hot cores and hot corinos, in cold dark clouds, and in cometary coma. Solid HCOOH is present in interstellar ices and in chondritic meteorites. 

\begin{figure}[t]
\begin{center}
 \includegraphics[width=2.4in]{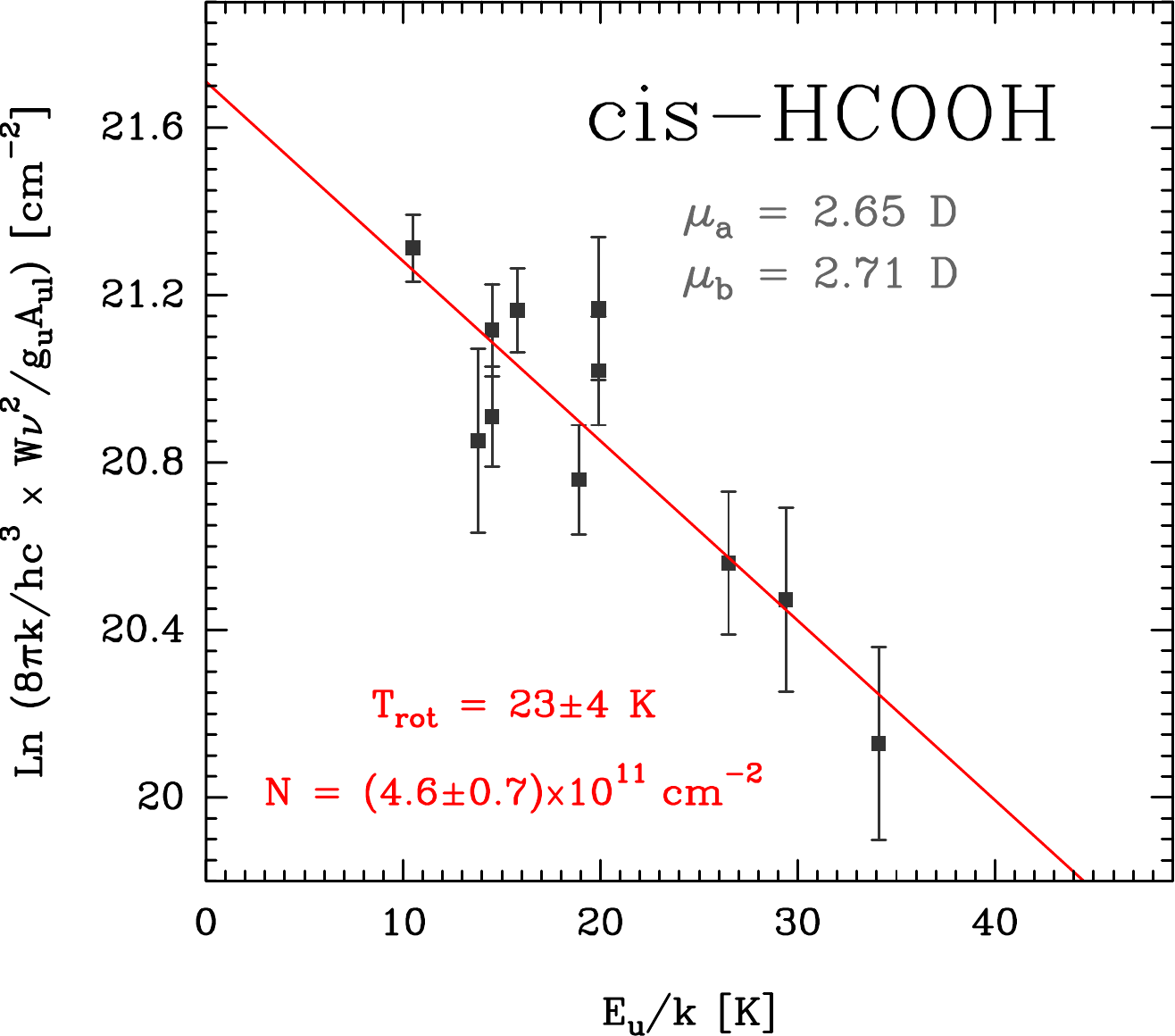} 
 \hspace*{0.6 cm}
 \includegraphics[width=2.1in]{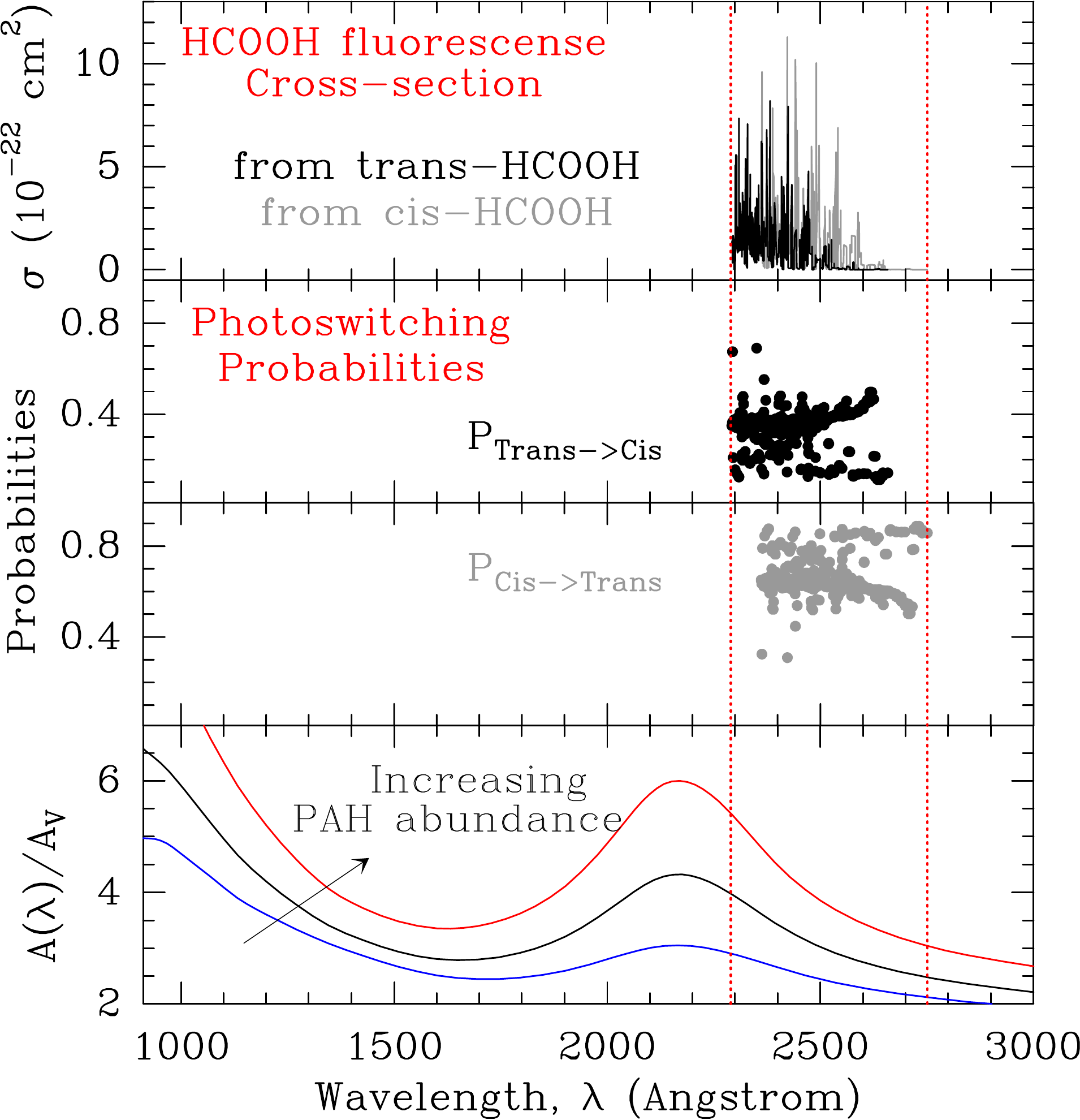} 
 \caption{ \textit{Left}) Rotational diagram of the \textit{cis}-HCOOH lines detected toward the Orion Bar.
 \textit{Right}) Top panel: \textit{Trans-} and \textit{cis}-HCOOH absorption cross-sections for photons with $E \lesssim$\,5\,eV (those leading to fluorescence) and photoisomerization probabilities. 
  Bottom panel: standard dust extinction curve (blue) and effect of an increased PAH abundance (\cite[Cuadrado et al. 2016]{Cuadrado_2016}).}
\label{fig5}
\end{center}
\end{figure}

The energy barrier to internal conversion from \textit{trans}- to \textit{cis}-HCOOH is much higher than the thermal energy available in molecular 
clouds\footnote{The ground-vibrational state of \textit{cis}-HCOOH is \mbox{1365 $\pm$ 30 cm$^{-1}$} higher in energy than that of the
\textit{trans} conformer. The energy barrier to internal rotation (the conversion from \textit{trans} to \textit{cis}) is
 approximately 4827\,cm$^{-1}$ ($\approx7000$\,K in temperature units).}.
Thus, only the most stable conformer (\textit{trans}) is expected to exist in detectable amounts. In Cuadrado et al. (\cite[2016]{Cuadrado_2016}) we reported the first interstellar detection of \textit{cis}-HCOOH.
Its presence in the Orion Bar (but not in the hot core of Orion nor in the
cold core of Barnard~1-b) with a  very low \mbox{\textit{trans}-to-\textit{cis}} abundance ratio of \mbox{2.8 $\pm$ 1.0} ($>$100 and $>$60 in the Orion hot core and in \mbox{B1-b} respectively), supports a photoswitching mechanism: a given conformer absorbs a FUV photon with energies around 5\,eV (precisely in the $\lambda$$\simeq$2300-2800\,\AA~range, higher energy photons dissociate the molecule) that radiatively excites the molecule to electronic states above the interconversion barrier. Subsequent fluorescent decay leaves the molecule in a different
conformer form. This mechanism, which we specifically studied with \textit{ab initio} quantum calculations (to compute the $\sim$5\,eV-photon absorption cross--sections and the
\textit{trans}-to-\textit{cis} and \textit{cis}-to-\textit{trans} photoswitching probabilities) was not considered in the ISM before, but likely induces structural changes of a variety of molecules in PDRs.
In Cuadrado et al. (\cite[2016]{Cuadrado_2016}) we used the \textit{Meudon} PDR code to estimate the flux
of FUV photons at different positons of the Bar.
The well-known ``2175\,\AA~bump'' of the dust extinction curve
(produced by the absorption of FUV photons by PAHs and small carbonaceous grains)
 reduces the number of HCOOH dissociating photons relative to those producing fluorescence (\mbox{Figure\,\ref{fig5}} \textit{right}). We determine that at a cloud depth of 
  $A_V\approx 2-3$\,mag (for the Orion Bar conditions), the \textit{cis} conformer should be detectable with a \textit{trans}-to-\textit{cis} abundance ratio of 3.5-4.1. These ratios are remarkably close to the observed values. 
  Therefore, both the detection of \mbox{\textit{cis}-HCOOH}, and
  the low \mbox{\textit{trans}-to-\textit{cis}} abundance ratio in the Bar  
  (much lower than in environments shielded from FUV radiation)
  would agree with the photoswitching scenario.

To summarize,  the presence of transient reactive ions such as CH$^+$, SH$^+$ or HOC$^+$ is an unambiguous indication of FUV-irradiated warm molecular gas.
The presence of COMs, however, is more widespread, and now  includes harsh environments such as  strongly FUV-irradiated gas
(where conformers can photoisomerize). COM formation reflects the complicated interplay between gas and grain  chemistry. Hence, more theory and laboratory experiments of the possible formation routes are needed. 
ALMA and soon JWST (for \mbox{non-polar} molecules, H$_2$ lines, PAHs and dust features) will open the study of
PDRs to sub-arcsecond resolution spectral-imaging. Exciting times ahead for PDR research.

\begin{thebibliography}{}

\bibitem[Black (1998)]{Black_1998} Black, J.~H.\ 1998, \textit{Faraday Discussions}, 109, 257 

\bibitem[Capak et al. (2015)]{Capak_2015} Capak, P.~L., Carilli, C., Jones, G., et al.\ 2015, \textit{Nature}, 522, 455 

\bibitem[Cernicharo et al.(2012)]{Cerni_2012} Cernicharo, J., Marcelino, N., Roueff, E., et al.\ 2012, \textit{ApJL}, 759, L43 

\bibitem[Champion et al. (2017)]{Champion_2017} 
Champion, J., Bern{\'e}, O., Vicente, S., et al.\ 2017, accepted in \textit{A\&A}, arXiv:1702.00251 

\bibitem[Cuadrado et al. (2015)]{Cuadrado_2015} Cuadrado, S., Goicoechea, J.~R., Pilleri, P., et al.\ 2015, \textit{A\&A}, 575, A82 

\bibitem[Cuadrado et al. (2016)]{Cuadrado_2016} Cuadrado, S., Goicoechea, J.~R., Roncero, O., et al.\ 2016, \textit{A\&A}, 596, L1 

\bibitem[Cuadrado et al. (2017)]{Cuadrado_2017} Cuadrado, S., Goicoechea, J.~R., Cernicharo, J., et al.\ 2017, \textit{A\&A},  603, A124 


\bibitem[Dalgarno \& McCray (1972)]{Dalgarno_1972} Dalgarno, A., \& McCray, R.~A.\ 1972, \textit{ARAA}, 10, 375 

\bibitem[Fuente et al.(2003)]{Fuente_2003} Fuente, A. et al. \ 2003, \textit{A\&A}, 406, 899 


\bibitem[Gerin et al. (2016)]{Gerin_2016} Gerin, M., Neufeld, D.~A., \& Goicoechea, J.~R.\ 2016, \textit{ARAA}, 54, 181 

\bibitem[Godard \& Cernicharo (2013)]{Godard_2013} Godard, B., \& Cernicharo, J.\ 2013, \textit{A\&A}, 550, A8 

\bibitem[Goicoechea \& Le Bourlot (2007)]{Goico_2007} Goicoechea, J.~R., \& Le Bourlot, J.\ 2007, \textit{A\&A}, 467, 1 

\bibitem[Goicoechea et al. (2015)]{Goico_2015} Goicoechea, J.~R., Teyssier, D., Etxaluze, M., et al.\ 2015, \textit{ApJ}, 812, 75 

\bibitem[Goicoechea et al. (2016)]{Goico_2016} Goicoechea, J.~R., Pety, J., Cuadrado, S., et al.\ 2016, \textit{Nature}, 537, 207 

\bibitem[Goicoechea et al. (2017)]{Goico_2017} Goicoechea, J.~R., Cuadrado, S., Pety, J., et al.\ 2017, \textit{A\&A}, 601, L9 

\bibitem[Guzm{\'a}n et al. 2014]{Vivi_2014} Guzm{\'a}n, V.~V., Pety, J., Gratier, P., et al.\ 2014, \textit{Faraday Discussions}, 168, 103 


\bibitem[Hollenbach \& Tielens (1999)]{Hollenbach_1999} Hollenbach, D.~J., \& Tielens, A.~G.~G.~M.\ 1999, \textit{Reviews of Modern Physics}, 71, 173 

\bibitem[Liszt et al. (2006)]{Liszt_2006} Liszt, H.~S., Lucas, R., \& Pety, J.\ 2006, \textit{A\&A}, 448, 253 

\bibitem[Nagy et al. (2013)]{Nagy_2013} Nagy, Z., Van der Tak, F.~F.~S., Ossenkopf, V., et al.\ 2013,  \textit{A\&A}, 550, A96 


\bibitem[Pety et al. (2012)]{Pety_2012} Pety, J., Gratier, P., Guzm{\'a}n, V., et al.\ 2012, \textit{A\&A}, 548, A68 

\bibitem[Russell et al. (1980)]{Russell_1980} Russell, R.~W., Melnick, G., Gull, G.~E., \& Harwit, M.\ 1980, \textit{ApJL}, 240, L99 


\bibitem[Walmsley et al. (2000)]{Walmsley_2000} Walmsley, C.~M., Natta, A., Oliva, E., \& Testi, L.\ 2000, \textit{A\&A}, 364, 301 


\bibitem[Zanchet et al. (2013)]{Zanchet_2013} Zanchet, A., Ag{\'u}ndez, M., Herrero, V.~J., Aguado, A., \& Roncero, O.\ 2013, \textit{AJ}, 146, 125 



\end{thebibliography}
\end{document}